\documentclass{aastex631}
\usepackage[utf8]{inputenc}
\usepackage{hyperref}
\usepackage{nameref}

\shorttitle{Persistent shift of the Crab pulsar}
\shortauthors{Zheng et al.}
\graphicspath{{./}{figures/}}
\usepackage{color}

\begin{document}

\title{The persistent shift in spin-down rate following the largest Crab pulsar glitch rules out external torque variations due to starquakes}

\author[0000-0001-8868-4619 ]{Xiao-Ping Zheng}
\affiliation{Institute of Astrophysics, Central China Normal University, Wuhan 430079, China\\}
\affiliation{Department of Astronomy, School of Physics, Huazhong University of Science and Technology, Wuhan 430074,  China\\}
\author[0000-0003-1473-5713]{Wei-Hua Wang}
\affiliation{Department of Physics, Wenzhou University, Wenzhou 325035, China \\}

\author[0000-0001-6406-1003]{Chun Huang}
\affiliation{Institute of Astrophysics, Central China Normal University, Wuhan 430079, China\\}
\affiliation{Physics Department and McDonnell Center for the Space Sciences, Washington University in St. Louis; MO, 63130, USA}

\author{Jian-Ping Yuan}
\affiliation{Xinjiang Astronomical Observatory, Chinese Academy of Sciences, Urumqi, Xinjiang 830011, China\\}
\author{Sheng-Jie Yuan}
\affiliation{Shenzheng Foreign Languages School,Guangdong 518083,China}
\email{zhxp@ccnu.edu.cn (XPZ)}\\
\email{wang-wh@wzu.edu.cn (WHW)}



\begin{abstract}

It was previously believed that, the long-term persistent increase in the spin-down rate of the Crab pulsar following a glitch is direct evidence of a starquake-induced glitch or at least related to a starquake. Using radio data covering 1710 days following the 2017 glitch, we obtain an extreme persistent increase of the spin-down rate, which allows to test two prevailing models related to starquake through an interrelation analysis between glitch size (the amplitude of the frequency step at a glitch) and persistent increase in the spin-down rate of the star. Our results do not support the hypothesis that glitches induce the external torque variation of the Crab pulsar, which may indicate no occurrence of starquake during the Crab pulsar glitch. This can explain why no changes in the radio and X-ray flux, pulse profile and spectrum of the Crab pulsar have been observed. We also suggest an internal mechanism due to superfluidity as an explanation for the long-term persistent shift in spin-down rate of the Crab pulsar following the relatively large glitches.
\end{abstract}



\section{Introduction}
\label{sec1}
Pulsars are generally extremely stable rotators and show regular spin-down trends.
However, the steady spin-down of many young pulsars is occasionally interrupted by the so-called glitch phenomenon, which corresponds to increases in rotational frequency and spin down rate of pulsars, followed by a recovery process to a new spin down rate within timescales of tens to hundreds of days~\footnote{The recovery timescale(s) can be found at the URL https://www.atnf.csiro.au/people/pulsar/psrcat/glitchTbl.html.}.
The fractional glitch size, defined as $\Delta\nu/\nu$, is in the range $\sim 10^{-11}-10^{-6}$, where $\nu$ is the pulsar rotation rate, $\Delta\nu$ is the step increase in rotational frequency.
Similarly, the fractional increase in spin down rate, $\Delta\dot\nu/\dot\nu$, ranges from $10^{-5}$ to $10^{-2}$, $\dot\nu$ is the pulsar spin down rate, $\Delta\dot\nu$ is the step increase in pulsar spin down rate, generally, $\Delta\dot\nu<0$.
Although nowadays there were a few sporadic cases possibly related to glitch activity, for example, the 2016 December glitch of the Vela pulsar~\citep{2018Natur.556..219P}, the 2007 glitch of the high magnetic field pulsar J1119-6127~\citep{2015MNRAS.449..933A} and the 2022 glitch of PSR J0742-2822~\citep{2024arXiv241217766Z}, no robust evidence for permanent surface changes on neutron stars has been found in rotation-powered pulsars (RPPs).
As of this writing, 682 glitches have been observed in 229 pulsars, among which are the prolific Crab pulsar and Vela pulsar~\footnote{https://www.jb.man.ac.uk/pulsar/glitches/gTable.html~\citep{2022MNRAS.510.4049B}.}.

Theoretically, there is currently no conclusive result on the physical origin of the glitch phenomenon, but two models are highly proposed, the starquake model~\citep{1969Natur.223..597R,1971AnPhy..66..816B,2014MNRAS.443.2705Z,2018MNRAS.476.3303L,2020MNRAS.491.1064G,2021A&A...654A..47R,2021MNRAS.500.5336W,2023MNRAS.523.3967L} and the superfluid model~\citep{1969Natur.224..673B,1975Natur.256...25A,1984ApJ...276..325A,1999PhRvL..83.3362L}.

In the starquake model, a glitch occurs when the accumulated stress of the material in the solid crust reaches its critical stress during the long-term spin down process of the pulsar~\citep{1969Natur.223..597R}. The critical stress is determined by the strength of the crustal material. On the basis of current neutron star models, starquake mechanism would lead to spin-ups of the magnitude observed in the Crab pulsar every few years, i.e., $\Delta\nu/\nu\sim 10^{-8}$, but the Vela quake of such magnitude $\Delta\nu/\nu\sim 10^{-6}$ may be repeated for several hundred years~\citep{1971AnPhy..66..816B,1972NPhS..235...43P}.

For the superfluid model, the interior superfluid component  of the neutron star (NS) rotates faster than the crystalline crust, the angular momentum of the interior superfluid component is contained in an array of superfluid vortices, and it decreases its rotation rate through vortex outward migration. Glitches arise from an exchange of angular momentum between the superfluid component of the star and the crystalline crust~\citep{1969Natur.224..673B,1975Natur.256...25A,2022ApJ...939....7H}. The following recovery process represents the response of vortex creep to glitch-induced rotation changes. During the steady state, when no glitch occurs, vortices do not migrate, which is called pinned to the crustal lattice sites, whereas the glitch process corresponds to vortices unpin from the pinning site. Because the recovery to pre-glitch spin-down state in Vela-like pulsars has been successfully described in the vortex creep model developed by Alpar et al. ~\citep{1984ApJ...276..325A}, currently the vortex model was thought to be the standard picture for pulsar glitches. Typically, Vela pulsar glitch includes several rapid exponential components in the spin down rate (the timescales range from days to tens of days) and a long-term linear increase in the spin-down rate~\citep{1990Natur.345..416F,1993ApJ...409..345A,2018IAUS..337..197M}.
In the vortex creep picture, weakly pinned regions have linear dynamical responses to crustal rotational changes and contribute the observed exponential recoveries in the spin-down rate, while the strongly pinned region has a non-linear dynamical response that results in a long-term linear increase in spin-down rate, as observed in Vela and other young pulsars~\citep{1993ApJ...409..345A}.

However, the post-glitch behavior in the Crab pulsar was quite different from that of the Vela pulsar. Large glitches in the Crab pulsar never recovered to the pre-glitch spin down rate~\citep{1992ApJ...390L..21L}. Three glitches in 1975 (MJD 42447.26(4)), 1989 (MJD 47767.504(3)), and 2011 (MJD 55875.5(1)) are classic events that will be called ``isolated" glitches in the next section. The post-glitch spin down rate of the Crab pulsar decreases rapidly within about 100 days after each glitch, and then it continues to increase in a quasi-exponential way toward a new stable value on a timescale of around 320 days (see Eq.(6) and figure in ~\cite{2015MNRAS.446..857L}). The difference between the new stable state and the pre-glitch one represents a permanent increase in spin-down rate of the star (persistent shift, $\Delta\dot\nu_{\rm p}$, hereafter). Meanwhile, by enough time after the glitch, the pulsar was spinning slower than the expected rates had the glitches not occurred, called frequency deffcits that was illustrated in figure 1 in ~\cite{1992ApJ...390L..21L}.

Owing to these post-glitch behaviors, the starquake is believed to be needed for the explanation of glitches in the Crab pulsar. The persistent shift might be explained by allowing changes in NS interior structure or surroundings of the NS~\citep{1992ApJ...390L..21L,1996ApJ...459..706A} due to starquake. There are two kinds of models accounting for persistent shift, global starquake~\citep{1969Natur.223..597R} or starquake-triggered decoupling of a portion of the stars superfluid interior from external torque~\citep{1993ApJ...409..345A,1996ApJ...459..706A}.
The former means that global starquake could decrease the moment of inertia (MOI) directly or the crust cracking due to either superffuid or spin-down stresses brings about a motion of the crustal plates toward the rotational axis of the star, producing a corresponding decrease in the moment of inertia~\citep{2018MNRAS.473..621A}.
The latter revealed that the so-called newly formed depletion region in the ``vortex capacitor" model also results in a permanent decrease in the effective NS MOI that the external torque acts on~\citep{1996ApJ...459..706A}. Although these models are constructed through essentially structural readjustment of the star to get a higher spin-down rate of NS, a decrease in NS MOI solely through structural readjustment cannot produce both the persistent shift and the frequency deficit observed in the Crab pulsar simultaneously~\citep{1992ApJ...390L..21L,1997ApJ...478L..91L}.
As decrease in NS MOI will increase the spin-down rate, but angular momentum conservation would require the star to spin more rapidly than had the glitch not occurred. This is obviously contrary to the observations of frequency deficits ~\citep{1992ApJ...390L..21L}: the persistent shifts in the spin-down rate following the 1975 and 1989 Crab glitches eventually caused the star to spin less rapidly than had the glitch not occurred~\citep{1997ApJ...478L..91L}.

Based on the above considerations, some models regarding an increased external torque seem promising to account for persistent shift.
Ruderman proposed that, NS crust cracking leads to ``plate tectonic" activity~\citep{1991ApJ...366..261R}.
Link et al. therefore argued that, forces exerted by pinned vortices on the crust would move crustal plates toward the equator, thus the magnetic dipole moment of the star becomes more misaligned with respect to the rotation axis, resulting in an increase in the external torque~\citep{1992ApJ...390L..21L}.
Using the persistent shift accumulated during 23 years (between 1969 and 1992), Link \& Epstein estimated that, the magnetic inclination angle should have increased by about 1/70 rad during the entire about 1000 years lifetime of the Crab pulsar.
In 1998, Link et al. proposed another possibility within starquake scenario.
They proposed that, a starquake perturbs the star's mass distribution due to a propagation of the surface material to higher latitudes along possible faults, producing a misalignment of the angular momentum and spin axes.
Subsequently, damped precession to a new rotational state increases the angle between the rotation and magnetic axes that results in an additional torque~\citep{1998ApJ...508..838L}.

The above two models are  apparently related to starquake. In this work, we try to test them through the observed persistent shifts of the Crab pulsar.
Our basic idea is presented here.
Lyne et al. pointed out that,  the absolute value of persistent shift following relatively large glitch in the Crab pulsar is approximately proportional to the glitch size~\citep{2015MNRAS.446..857L}.
The Crab pulsar experienced its largest glitch ever observed in 2017, with $\Delta\nu/\nu=0.516\times 10^{-6}$ and $\Delta\nu=1.530\times 10^{-5}~{\rm {Hz}}$~\citep{2018MNRAS.478.3832S}.
This large glitch is followed by several small glitches~\citep{2022MNRAS.510.4049B}.
After all these years, all the post-glitch transient components in pulsar spin frequency and spin down rate have decayed, therefore, we have a chance to measure its persistent shift and reanalyze  the $\left |\Delta\dot\nu_{\rm p}\right |-\Delta\nu$ relation for the Crab pulsar.
We derive their expected $\left |\Delta\dot\nu_{\rm p}\right |-\Delta\nu$ relation(s) based on the two models respectively~\citep{1991ApJ...366..261R,1992ApJ...390L..21L,1998ApJ...508..838L,2000ASSL..254...95E}.
Comparing the predications based on the models with measured results based on observed data including the 2017 glitch, we test if the persistent shift is caused by starquake induced external torque variation.

This manuscript is organized as follows. We first measure the persistent shift caused by the 2017 Crab pulsar glitch in Section \ref{sec2}.
In Section \ref{sec3},  we statistically reanalyze the 11 shift values of 30 glitch events for over 50 years. We give a new fitting function of $|\Delta\dot\nu_{\rm p}|-\Delta\nu$ relation and calculate the significance level of deviations from linear correlation as proposed by Lyne et al. ~\citep{2015MNRAS.446..857L} for the large Crab pulsar glitch in 2017.
In Section \ref{sec4}, we apply the prevailing models based on external torque variations to derive the model predicted $\left |\Delta\dot\nu_{\rm p}\right |-\Delta\nu$ relations and compare the theoretical expectations with our data fittings in Section 3.
Finally, conclusions and discussions are presented in Section \ref{sec5}.

\section{The spin-down offset following Crab glitches within the year 2017 and 2019}
\label{sec2}

Currently, 30 glitches have been observed from the Crab pulsar between 1968 and 2024~\citep{2022MNRAS.510.4049B}.
However, ever since the first discovery of persistent shift following Crab pulsar glitch in 1975~\citep{1977AJ.....82..309G,1981A&AS...44....1L}, only ten persistent shifts have been measured, they are summarized in Lyne et al.~\citep{2014MNRAS.440.2755E,2015MNRAS.446..857L}.
The reason for the relatively small number of persistent shift values is that, the glitches whose persistent shift can be measured should be isolated, namely, each having no other detectable glitches within 800 days before or 1200 days after the epoch of the glitch.
Unfortunately, the post-glitch relaxation process is often contaminated by the occurrence of nearby glitches.
Only five out of 30 are isolated or relatively isolated glitch events.
The persistent shifts after these five glitches are notable and explicit, as shown in Figure 3 and Table 3 in ~\cite{2015MNRAS.446..857L}.
As for the other five persistent shift values, they correspond to the cumulative effects of two or three neighboring glitches, the shift of the individual glitch is unknown.
For example, the persistent shift value of $-116(5)\times10^{-15}~{\rm Hz/s}$ corresponds to the cumulative effect of the two glitches on MJD 50260.031(4) and MJD 50458.94(3)~\citep{2015MNRAS.446..857L}.

The 2017 November (MJD 58064.555) glitch is discovered to have the largest size in the Crab pulsar to date~\citep{2018MNRAS.478.3832S} and should display the
largest persistent shift value according to the linear fitting to $\Delta\nu-\Delta\dot\nu_{\rm p}$ relation in~\cite{2015MNRAS.446..857L}.
However, measurement of its persistent shift is obscured by the occurrence of the following three much smaller glitches in about 600 days, namely, 2018 March (MJD 58237.357 $\Delta\nu/\nu=4.08(22)\times 10^{-9}$~\citep{2020MNRAS.491.3182B}), 2018 November (MJD 58470.7, $\Delta\nu/\nu=2.3(6)\times 10^{-9}$~\citep{2022MNRAS.510.4049B}) and 2019 July (MJD 58687.565, $\Delta\nu/\nu=31.7(12)\times 10^{-9}$~\citep{2021MNRAS.505L...6S}) glitches.
So we can only obtain the persistent shift value that corresponds to the cumulative effect of the above four glitches.

\begin{figure}[ht!]
\centering
\includegraphics[width=0.5\textwidth,height=0.3\textwidth]{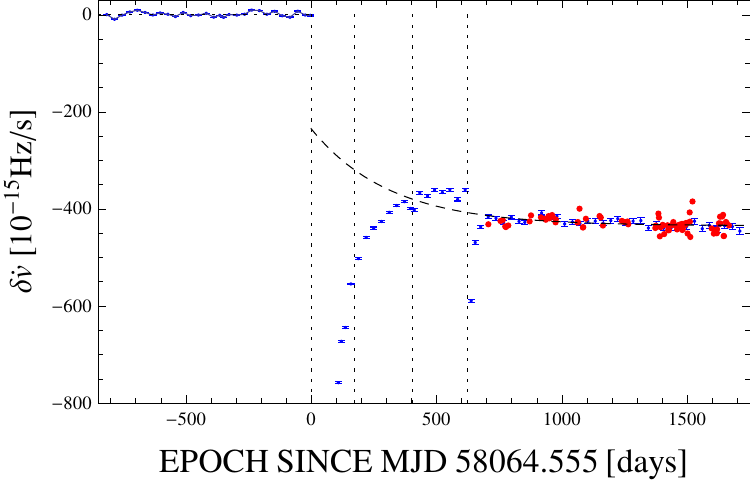}
\caption{Frequency derivative residual $\delta\dot\nu$ plotted against epoch since MJD 57249.0. Blue dots: data taken from the Jodrell Bank Crab pulsar monthly ephemeris. Red dots: the data from Nanshan 26~m telescope. The first vertical dotted line in the left corresponds to the epoch of 2017 Crab pulsar glitch on MJD 58064.555. The other three vertical lines represent the epochs of glitch on MJD 58237.357, MJD 58470.7, and MJD 58687.565 respectively. The dashed curve after the glitch on MJD 58064.555 represents the best fit of the exponential decay model, the red dots were not used for the fitting.}
\label{fig1}
\end{figure}

Our procedure is as follows.
Firstly, we obtain the frequency derivative $\dot\nu$ values of the Crab pulsar. The data are collected from the Jodrell Bank Crab pulsar monthly ephemeris~\footnote{https://www.jb.man.ac.uk/pulsar/crab.html}~\citep{2022MNRAS.510.4049B}.
We have acquired the frequency derivative data ranging from MJD 57249 to MJD 59775, where the first point is over 800 days preceding the glitch on MJD 58064.555.
Totally, 93 values of  $\dot\nu$ are included in this analysis, 29 ones preceding this glitch and 64 ones after the glitch.
Secondly, variation in long-term rotation frequency with time is characterized as a Taylor series of frequency derivative of the form
\begin{equation}
\nu(t)=\nu_{0}+\dot\nu_{0}(t-t_{0})+\frac{1}{2}\ddot\nu_{0}(t-t_{0})^{2}+\frac{1}{6}\dddot\nu_{0}(t-t_{0})^{3}+\delta\nu,
\end{equation}
where $\dot\nu_{0}$, $\ddot\nu_{0}$, and $\dddot\nu_{0}$ represent the first, the second, and the third derivatives of rotation frequency respectively, $\delta\nu$ is the residual in rotation frequency, $t_{0}$ is the reference time.
We here prefer to focus on the characteristics of the frequency derivatives over time. So, likewise, the variation in long-term  spin down rate with time can be expressed in the form
\begin{equation}
\dot\nu(t)=\dot\nu_{0}+\ddot\nu_{0}(t-t_{0})+\frac{1}{2}\dddot\nu_{0}(t-t_{0})^{2}+\delta\dot{\nu}.
\label{departure}
\end{equation}
where $\delta\dot{\nu}$ denotes frequency derivative residual. In our case, the term including $\dddot\nu_{0}$ in Eq.(2) could be neglected, as $\dddot\nu_{0}$ is much too small and we are discussing the short-term spin down rate behavior around the glitch (several years after the glitch). Therefore,
fitting the linear portion in Eq.(\ref{departure}) to the 29 $\dot\nu$ before the glitch on MJD 58064.555 as a function of time with Eq.(\ref{departure}), the corresponding slope $\ddot\nu_0$ is $1.174(4)\times 10^{-20}~{\rm Hz/s^{2}}$ and the spin down rate at the glitch epoch is $-368611(2)\times 10^{-15}~{\rm Hz/s}$, where $t_{0}={\rm MJD}~57249.0$.
Thirdly, subtracting the straight line in Eq.(\ref{departure}) from the above 93 $\dot\nu$ results in the derivative residuals,$\delta\dot\nu$, shown as blue dots in Fig.\ref{fig1}.
Our Fig.\ref{fig1} is similar to Fig.3 in ~\cite{2015MNRAS.446..857L} and Fig.1 in~\cite{2017A&A...597L...9V}
in terms of processing methodology. Note that we have also determined a series of $\dot\nu$ values from fits to 5-20-day data sets of Nanshan $26~{\rm m}$ telescope using TEMPO2~\citep{2006MNRAS.369..655H}, we then derived the corresponding $\delta\dot\nu$ and plotted them as red points in Fig.\ref{fig1} for comparison.
The Nanshan data span  from MJD 58771.5 to MJD 59731.3.
The trends of the blue and red dots are consistent.

After completing the procedures outlined above, we can finally measure the persistent shift value.
On the one hand, as we can see from Fig.\ref{fig1}, the recovery process of the glitch on MJD 58064.555 is interrupted by the three nearby glitches, so we can only measure the cumulative effect of the four glitches.
On the other hand, as stated in paragraph 5 in Section \ref{sec1} and the first paragraph of this section, this glitch is not isolated, so $\delta\dot\nu$ following this glitch can not be fully fitted by the exponential decaying function in Eq.(6) in ~\cite{2015MNRAS.446..857L} in principle.
However,  the persistent shift can be determined by calculating the limit, $\lim\limits_{t \to +\infty} \delta\dot\nu=\Delta\dot\nu_{\rm p}$. This hence implies that the exponential decay would be fitted to the data points when the recovery time is long enough.
In our case, fitting the data ranging from MJD 58832 to MJD 59775 with $\delta\dot\nu=\Delta\dot\nu_{\rm p}\times[\kappa\times {\rm exp}(-t/\tau)-1.0]$, we have $\Delta\dot\nu_{\rm p}=-434(3)\times 10^{-15}~{\rm Hz/s}$ and $\tau=312.3\pm 33.8~{\rm days}$ when $\kappa=0.46$ is taken as proposed in ~\cite{2015MNRAS.446..857L}. Interestingly, the timescale $\tau$ determined by fitting the last points is  almost consistent with the time scale derived from $320\pm 20~{\rm days}$ from those isolated events presented in~\cite{2015MNRAS.446..857L}.
 Thus, the fitting function is compared with the preglitch value, shown in Fig.\ref{fig1}, to obtain the largest persistent shift so far, $\Delta\dot\nu_{\rm p}=-434(3)\times 10^{-15}~{\rm Hz/s}$.
Although the result is contributed by all 4 glitches, the glitch occurred in November 2017 contributed the most since its glitch size amounts to $93\%$ of the cumulated glitch sizes of the 4 glitches, therefore, $\Delta\dot\nu_{\rm p}=-434(3)\times 10^{-15}~{\rm Hz/s}$ can be viewed as the upper limit of the persistent shift caused by the November 2017 glitch.
We also noticed that, when comparing the three glitches following the November 2017 glitch with previous Crab pulsar glitches, the total glitch size of the three glitches is similar to that of the two successive glitches on MJD 50260.031 and 50458.94 (see our Table 1).
If we regard the three glitches after November 2017 glitch and the two successive glitches on MJD 50260.031 and 50458.94 as equivalent in terms of persistent shift, by subtracting the cumulated persistent shift of the latter, we get the lower limit of persistent shift of the 2017 November glitch to be $\Delta\dot\nu_{\rm p}\simeq -318(3)\times 10^{-15}~{\rm Hz/s}$.
We present all the 11 persistent shifts in Table \ref{table1} together with previously measured shifts.

\begin{deluxetable*}{lccccc}
\tabletypesize{\scriptsize}
\tablewidth{0pt}
\tablecaption{11 persistent shift measurements after Crab glitches}
\tablehead{
\colhead{Glitch} & \colhead{MJD}&\colhead{Glitch Size}& \colhead{Persistent Shift}  \\
\colhead{} &\colhead{Date} &\colhead{${\rm\mu}$Hz} & \colhead{($10^{-15}{\rm Hz/s}$)}
\label{table1}
}
\startdata
 1975 February    & 42447.26(4)      & 1.08(1)     & $-112(2)$\\
 1989 August      & 47767.504(3)     & 2.43(1)     & $-150(5)$\\
 1996 June        & 50260.031(4)     & 0.953(4)    &           \\
 1997 January     & 50458.94(3)      & 0.18(1)     &$-116(5)$$^a$  \\
 1999 October     & 51452.02(1)      & 0.20(1)     & $-25(3)$\\
 2000 July        & 51740.656(2)     & 0.75(1)     & \\
 2000 September   & 51804.75(2)      & 0.105(3)    &$-53(3)$$^a$\\
 2001 June        & 52084.072(1)     & 0.675(3)    & \\
 2001 October     & 52146.7580(3)    & 0.265(1)    & $-70(10)$$^a$\\
 2002 August      & 52498.257(2)     & 0.101(2)    & \\
 2002 September   & 52587.20(1)      & 0.050(3)    &$-8(2)$$^a$\\
 2004 March       & 53067.0780(2)    & 6.37(2)     & \\
 2004 September   & 53254.109(2)     & 0.145(3)    &\\
 2004 November    & 53331.17(1)      & 0.08(1)     &$-250(20)$$^b$\\
 2006 August      & 53970.1900(3)    & 0.65(1)     & $-30(5)$\\
 2011 November    & 55875.5(1)       & 1.18(2)     & $-132(5)$\\
 2017 March       & 57839.8(1)       & 0.067(2)    &\\
 2017 November    & 58064.555(3)     & 15.491(3)   & \\
 2018 March       & 58237.357(5) 	& 0.121(7)    &\\
 2018 November   & 58470.7(2)       & 0.070(8)    &\\
 2019 July        & 58687.565(4)	    & 0.938(4)	 &$-434(3)$$^c$
\enddata
\tablecomments{The first 10 sets in this table is derived from~\citep{2014MNRAS.440.2755E,2015MNRAS.446..857L}. The glitch data during 2017-2019 are taken from ~\citep{2018MNRAS.478.3832S,2020MNRAS.491.3182B,2021MNRAS.505L...6S,2022MNRAS.510.4049B}. $^a$ Incorporates the persistent shift of the previous glitch. $^b$ Incorporates the persistent shifts of the previous two glitches. $^c$ Incorporates the persistent shifts of the previous three glitches.}
\end{deluxetable*}

\section{Data analysis of correlation}
\label{sec3}

\begin{figure}[ht!]
\centering
\includegraphics[width=0.5\textwidth,height=0.3\textwidth]{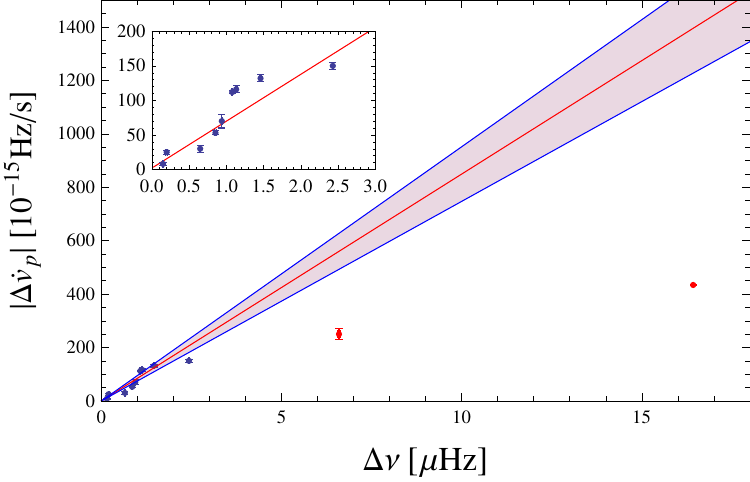}
\caption{The persistent shifts in spin-down rate vs glitch sizes. The red line denotes the fitting to the data dots that do not include the 2004 and 2007 glitches (denoted by red dots). The inner 68\% con?dence interval are shown by shaded region between the blue lines. The 2004 and 2017 glitches involved remarkable departure from the line and the 2017 is an extreme outlier. The inset shows the strong linear correlation for the small glitch events. }
\label{shift-size-linear}
\end{figure}

The correlation between $\Delta\nu$ and $\Delta\dot\nu_{\rm p}$ is re-analyzed in this work. Previously,
Lyne et al. has investigated the inter-dependence of $\Delta\dot\nu_{\rm p}$ and $\Delta\nu$ according to the results of previous 24 glitches till 2011, giving linear relationship as follows~\citep{2015MNRAS.446..857L},
\begin{equation}\label{relation 1}
\left |{\Delta\dot{\nu}_{\rm p}}\right | =7.0\times 10^{-8}\Delta\nu~{\rm Hz}~ {\rm s}^{-1}.
\end{equation}
This linear relationship looks like a good match to data points on the logarithmic plot (see figure 5 in~\cite{2015MNRAS.446..857L}), but we noticed that, $\left |\Delta\dot\nu_{\rm p}\right |$ is actually loosely related to $\Delta\nu$ as the large glitch in 2004 resulted in a $\left |\Delta\dot\nu_{\rm p}\right |$ departed from their linear relation apparently, which is easier to be shown if we use uniform coordinates in place of logarithmic ones.
Therefore, we put the data points from Table 1 in the $\left |\Delta\dot\nu_{\rm p}\right |-\Delta\nu$ graph by letting $\Delta\nu$ and $\left |\Delta\dot\nu_{\rm p}\right |$ be the horizontal and vertical coordinates, respectively.
Neglecting the two very large events, linear fitting to the data  gives
 \begin{equation}\label{relation 2}
 \left |{\Delta\dot{\nu}_{\rm p}}\right | =8.5(10)\times 10^{-8}\Delta\nu~{\rm Hz}~ {\rm s}^{-1}.
 \end{equation}
The two fittings are almost consistent although previous fitting covered all the events of that time, including the large glitch in 2004, implying that small events dominate the linear correlation.
The result of Eq.(\ref{relation 2}) is depicted in Figure~\ref{shift-size-linear}, the predicted values for the 2004 and 2017 events according to Eq.~(\ref{relation 2}) and the significance of their deviations from the measured $\Delta\dot\nu_{\rm p}$ are listed in Table 2.
Figure \ref{shift-size-linear} also shows that the data points of March 2004 and November 2017 events have departed from our linear fitting remarkably. Especially,
for the 2017 November glitch, the measured data point is a clear outlier from our linear trend.

\begin{deluxetable*}{lccccc}
\tabletypesize{\scriptsize}
\tablewidth{0pt}
\tablecaption{Two oversized shifts and deviations from linear correlation }
\tablehead{
\colhead{Glitch} & \colhead{Glitch Size}&\colhead{Measured shift}& \colhead{Predicted shift} & \colhead{68\% confidence interval($\sigma$)} & \colhead{Deviation significance}\\
\colhead{}  &\colhead{${\rm\mu}$Hz} & \colhead{($10^{-15}$ Hz/s)} & \colhead{($10^{-15}{\rm Hz/s}$)} & \colhead{($10^{-15}{\rm Hz/s}$)} & \colhead{($\sigma$)}
\label{table2}
}
\startdata
 2004 March       & 6.37(2)   &\\
 2004 September   & 0.145(3)  &\\
 2004 November    & 0.08(1)   &$-250(20)^b$ & -559.95 & 67.40 & 4.5\\
 2017 November    & 15.491(3) &\\
 2018 March       & 0.121(7)  &\\
 2018 November    & 0.070(8)  &\\
 2019 July        & 0.938(4)	 &$-434(3)^c$ & -1394.81 & 167.89 & 5.7
\enddata
\tablecomments{ $^b$ and $^c$, same as that indicated in Table\ref{table1}}
\end{deluxetable*}

\begin{figure}[ht!]
\centering
\includegraphics[width=0.5\textwidth,height=0.3\textwidth]{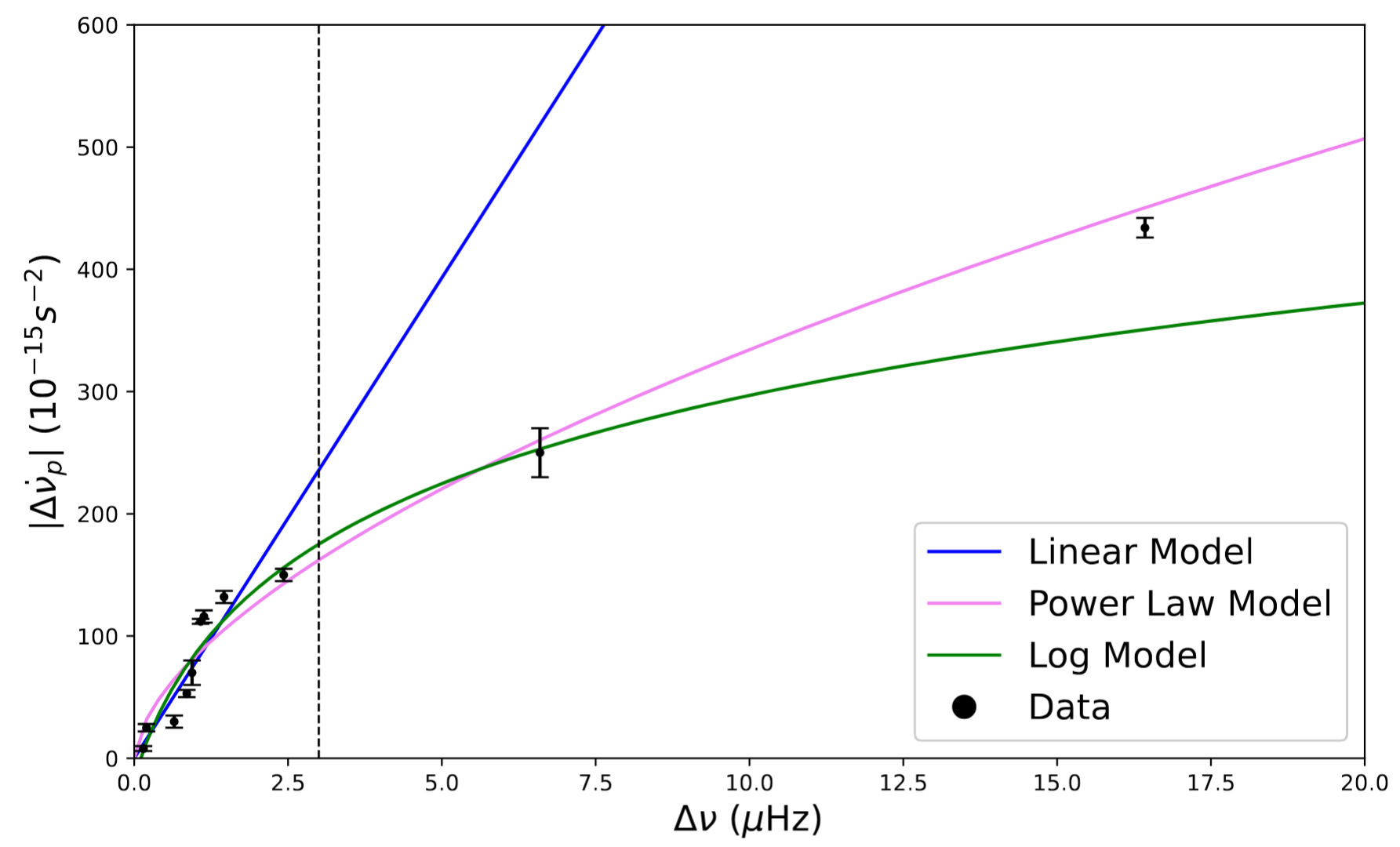}
\caption{Linear, power law, and logarithmic fittings to $\Delta\nu-|\Delta\dot\nu_{\rm p}|$, the fitting functions are Eq.(\ref{relation 2}), Eq.(\ref{power-law fitting}), and Eq.(\ref{logarithmic fitting}) respectively. The linear model was fitted only up to $\Delta\nu= 3~\mu$Hz, indicated by the dashed vertical line.}
\label{shift-size-nonlinear}
\end{figure}

Based on the above discussions, a linear $\left |\Delta\dot\nu_{\rm p}\right |-\Delta\nu$ relation is actually disfavored, nonlinear fittings are possibly needed.
We try to fit all data points with power-law and logarithmic functions, the mathematical fitting functions are
\begin{equation}\label{power-law fitting}
\left |{\Delta\dot{\nu}_{\rm p}}\right | =8.35\times 10^{-14}\Delta\nu^{0.60}~{\rm Hz}~ {\rm s}^{-1}
\end{equation}
and
\begin{equation}\label{logarithmic fitting}
\left |{\Delta\dot{\nu}_{\rm p}}\right | =1.15\times 10^{-13}\ln(0.87+1.25\Delta\nu)~{\rm Hz}~ {\rm s}^{-1}
\end{equation}
respectively, with goodness-of-fit ($R^2$- test) of 0.9681 and 0.9786.
The nonlinear fittings are shown in Figure \ref{shift-size-nonlinear}.
The two fittings depart greatly from the linear relationship represented by Eq.(\ref{relation 2}), but they match well with each other, as the power law fitting is the lead-order approximation of the logarithmic case, but the dominating contribution to the fitting function as evidenced by a slight difference of two goodness-of-fits, $R^2$.

\section{Tests of models regarding starquake-induced changes in external torque}
\label{sec4}
\begin{figure}[ht!]
\centering
\includegraphics[width=0.5\textwidth,height=0.3\textwidth]{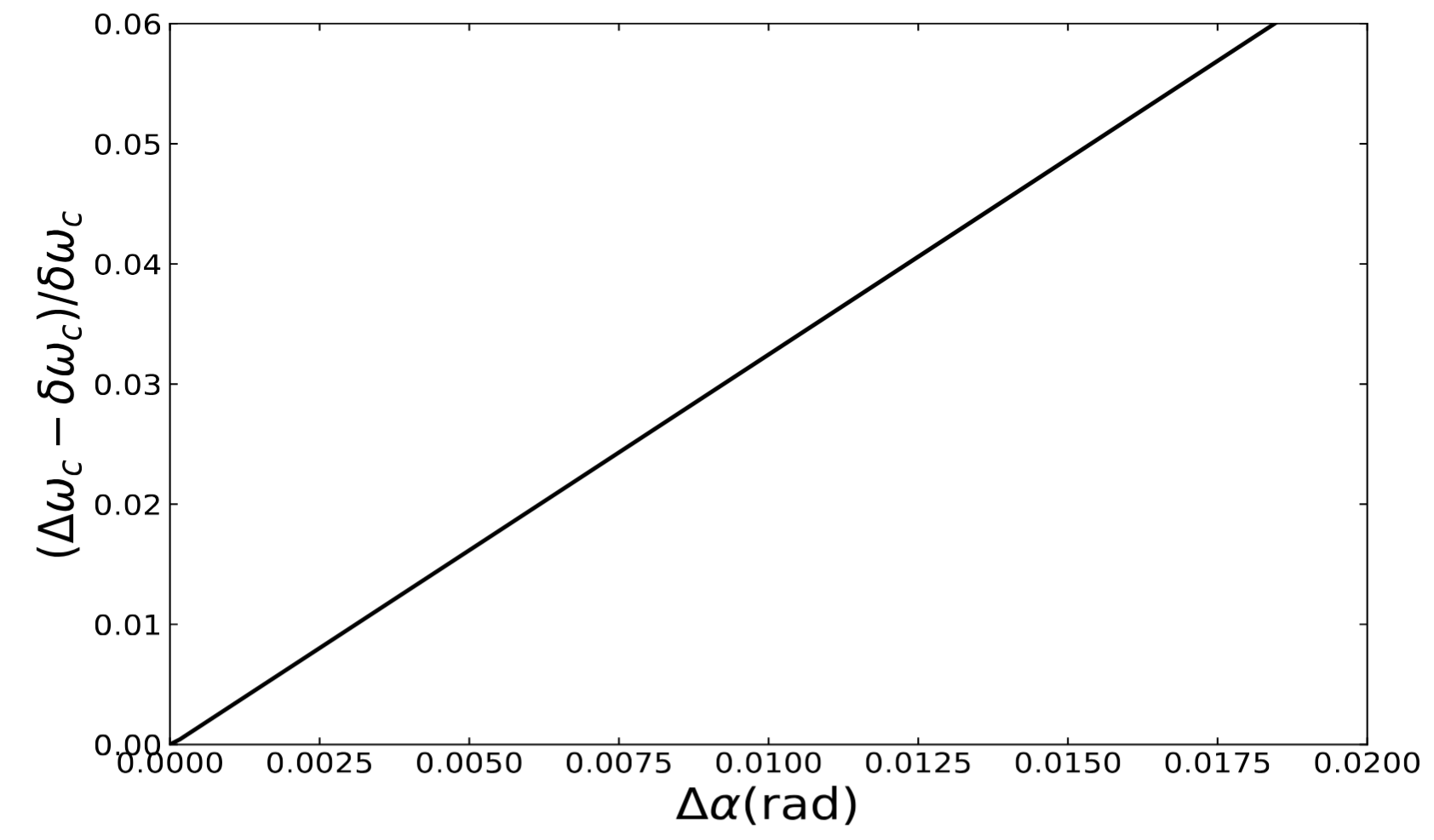}
\caption{The fractional deviations of critical lag change from linear relation with changes in inclination angle. Eq(\ref{eq6}) is a good approximation until $\Delta\alpha\sim 10^{-2}$ rad because the deviation of realistic $\Delta\omega_c(\Delta\alpha)$ from linear $\delta\omega_c(\Delta\alpha)$  is only $3\%$ at  $\Delta\alpha\sim 10^{-2}$ rad.}
\label{fig4}
\end{figure}

Theoretically, there are two prevailing models involving starquake or crust cracking as described in paragraph 7 in Section \ref{sec1}, which propose increases in the magnetic inclination angle via crust plate moving towards equator of the star or mass transfer to high latitude.
The first model is based on ``plate tectonic" activity proposed by Ruderman~\citep{1991ApJ...366..261R}, we call it ``tectonic model" hereafter.
In this model, the critical lag between angular velocity of the superfluid component pinned to the crust and the crustal lattice, indicating the lag value which would cause vortex line unpinning from the lattice, is seen as a function of the angle between the pinned position and rotational axis of the star, which is formulated as
 \begin{equation}\label{eq5}
\omega_{\rm c} ={ {\hbar R_N^3}\over{\pi R\sin\alpha f m_n b_z^3 }},
 \end{equation}
 where $\hbar$ is the Plank constant, $R_N$ the radius of the lattice nuclei, $R$ the radius of the star, $f$ the dimensionless factor depending on the amount of neutron superfluid in pinning region, $m_n$ the neutron mass, and $b_z$ the separation between nuclei.  As done in references~\citep{1992ApJ...390L..21L,1997ApJ...478L..91L}  the angle between a pinned position and the rotational axis,$\alpha$, is equal to the angle between the magnetic and rotational axes. Actually, the two angles are equal due to the balances in forces and torques. This can be proved as follows: If we assume that the lag between angular velocity of the superfluid component pinned to the crust and the crustal lattice, the angle between a pinned position and the rotational axis,the superfluid neutron density and the distance between pinning nuclei are respectively $\omega, \alpha', \rho_n$ and $b$, the magnitude of the force on a pinning nucleus because
of neutron superfluid flow past the pinned vortex line is $F_N=\omega R\sin\alpha' f\pi \rho_n b\hbar/m_n\equiv F_N^0\sin\alpha'$ (see Eq(23) in~\cite{1991ApJ...366..261R}). Accordingly, the torque on the crust of the star is $M_N=F_N^0R\sin^2\alpha'$. On the other hand, we can take the average external torque acting on a pinning nucleus as $M_B=B^2R^6\sin^2\alpha\Omega^3/(N6c^2)\equiv F_B^0R\sin^2\alpha$, with $F_B^0=B^2R^5\Omega^3/(N6c^2)$ for magnetic dipole model $\dot{\Omega}=-B^2R^6\sin^2\alpha\Omega^3/(6Ic^2)$, $B,\alpha, N,I,c$ and $\Omega$ represent the magnetic field, the magnetic inclination angle, the number of pinning nucleus,  the total moment of inertia of the star, the speed of light in vacuum and angular speed of the star respectively. Of course, the according force on the crust of the star is $ F_B\equiv F_B^0\sin\alpha$. For the pinning nucleus, the balances in forces and torques should be satisfied, i.e., $F_N=F_B$ and $M_N=M_B$. We thus have $\alpha'=\alpha$ with $F_N^0=F_B^0$.
During pulsar glitches, Eq.(\ref{eq5}) means that the changes in crustal rotational frequency, equal to changes in critical lag frequency, cause misaligned angles $\Delta\alpha$ that lead to external torque variations.
Solving $\Delta\omega_{\rm c}= \omega_{\rm c}(\alpha+\Delta\alpha)-\omega_{\rm c}(\alpha)$ according to Eq. (\ref{eq5}), we obtain the dependence of $ \Delta\omega_{\rm c} $ on $\Delta\alpha$, i.e., the function $\Delta\omega_{\rm c}(\Delta\alpha)$.
 When $\Delta\alpha$ reaches an infinitely small value, replacing $\Delta\alpha$ by $\delta\alpha$,  differentiating the function $\omega_{\rm c}(\alpha)$ from Eq.(\ref{eq5}), we have
 \begin{equation}
\delta\omega_{\rm c} =-\omega_{\rm c}{{\delta\alpha}\over{\tan\alpha}}.
\label{eq6}
 \end{equation}
Over what range of parameters does it work? We can get an answer from the relative fractional deviations denoted by ${\Delta\omega_c-\delta\omega_c}\over{\delta\omega_c}$. For magnetic dipole model, $\dot{\Omega}=-B^2R^6\sin^2\alpha\Omega^3/(6Ic^2)$, we have
\begin{equation}\label{eq7}
{{\Delta\dot{\nu}_{\rm p}}/{\dot{\nu}_0}}={{2\Delta\alpha}/{\tan\alpha}}.
\end{equation}
Here should be emphasized that the right-hand side in the equality could be different expressions of $\alpha$-dependence~\citep{2016ApJ...825...18N,2018MNRAS.478L..24K} but not affect our discussions. We calculate the shifts from Eq.(\ref{eq7}) for given $\alpha$ and $\Delta\alpha$. Figure \ref{fig4} shows the fractional deviations of the exact results from linear function of Eq.(\ref{eq6}). We find that the relative deviation is only $3\%$ until  $\Delta\alpha\sim 10^{-2}$ rad and extreme persistent shifts range from $10^{-10}$ to $10^{-12}$ Hz s$^{-1}$ for inclination angle $\alpha$ from $ 5^\circ$ to $ 89^\circ $. Based on Eq(\ref{eq7}) again, the largest measured shift of the Crab pulsar (see Table 1) is much smaller than that at $\Delta\alpha\sim 10^{-2}$ rad, thus a linear relationship between $ \Delta\omega_c $ and $\Delta\alpha$ always satisfies for the Crab pulsar.
The angular velocity of the superfluid component remains almost unchanged at the moment when a glitch causes a perturbation in the crust of the star, so we have $\delta\omega_c=2\pi\Delta\nu$. Combine Eq.(\ref{eq7}), Eq.(\ref{eq6}) becomes
\begin{equation}
\left |{\Delta\dot{\nu}_{\rm p}}\right | ={2\left | 2\pi\dot{\nu}_0\right |\over\omega _{\rm c}}\Delta\nu
\label{eq8}
\end{equation}
where $\dot{\nu}_0=-3.68\times 10^{-10}~\rm Hz/s$.  This equation can reproduce the fitting function of Eq.(\ref{relation 2})  for $ \omega_{\rm c} \sim5.4\times 10^{-2}$~{\rm{rad/s}}. However, the analysis based on Fig 2, Table 2 and Fig 3 imply that the observation of 2017 glitch strongly departures from the linear relationship (Eq.\ref{eq8}) and hence rules out the possibility of ``plate tectonic" activity.

\begin{figure}[ht!]
\centering
\includegraphics[width=0.5\textwidth,height=0.3\textwidth]{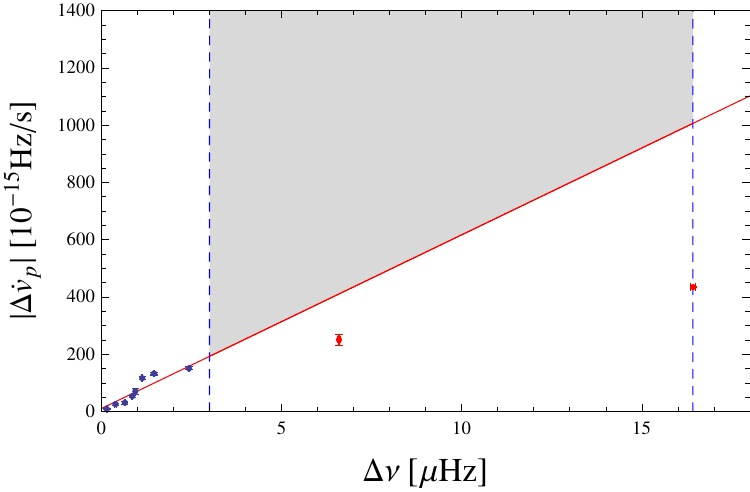}
\caption{Comparison of fault model to observed data. The red line is a lower limit and the shadow region presents an allowed window for the persistent shifts caused by glitches. The left vertical dashed line indicates $\Delta\nu=3~\mu$Hz inferred from Fig.\ref{shift-size-linear} and the right one at 16.62~$\mu$Hz is the accumulated sizes of glitches on MJD 58064.555. MJD 58237.357, MJD 58470.7, and MJD 58687.565.}
\label{fig5}
\end{figure}

For the second model (``fault model"hereafter), starquake may trigger pulsar glitches and result in jumps in spin-down rate simultaneously~\citep{1998ApJ...508..838L,2000ApJ...543..987F}.
Matter can slide
to higher latitudes along
faults. The creation of new principal axis of inertia due to mass redistribution will induce the shift of magnetic poles to the the equator by an angle (i.e. the increase in magnetic inclination angle) $\Delta\alpha=2.5\times 10^4\Omega_2^{-2}\delta I/I_{\rm crust}$~\citep{2000ASSL..254...95E}, where $\delta I$ denotes the moment of inertia excess created when starquake shifts the stellar matter asymmetrically and $I_{\rm crust}$ is crustal moments of inertia of the star. On the one hand, increases of tilt angle will induce a persistent shift according to Eq.(\ref{eq7}).
On the other hand, the processes relaxing to the new principal axis produces a rapid slowing of the super?uid component as well as a spin up of the crust~\citep{2000ASSL..254...95E}, the spin up magnitude can be expressed as
\begin{equation}
{\Delta\Omega\over\Omega}\leq -{I_{\rm unpinned}\over I}{\Delta\Omega_{\rm s}\over\Omega}={I_{\rm unpinned}\over I}\Delta\alpha,
\label{eq9}
\end{equation}
where $I_{\rm unpinned}$ and $I$ denote the moment of inertia of rapid vortex creep region and the total NS moment of inertia respectively, $\Delta\Omega_{\rm s}$ is the angular velocity change of the superfluid component. The $``<"$ sign occurs as the vortex creep would stop before the vortex lines move a full distance $\Delta\alpha R$ without enough lag velocity of pre-starquake, where $R$ is radius of the star.
Substituting Eq.(\ref{eq9}) into Eq.(\ref{eq7}) gives
\begin{equation}
\left |{\Delta\dot{\nu}_{\rm p}}\right | \geq {2\left |\dot{\nu}_0\right |\over\nu\tan\alpha}{I\over I_{\rm unpinned}}\Delta\nu.
\label{eq10}
\end{equation}
 The lower limit of the Eq.(\ref{eq10}) is  a linear function of $\Delta\nu$ which matches well with Eq.(\ref{relation 2})  for the parameters $\alpha\gtrsim 75^\circ$ and $I_{\rm unpinned}\lesssim 0.5~I_{\rm crust}$.
Eq.(\ref{eq10}) also means that, persistent shifts of large glitches should be above the linear relationship given by the fitting to $\Delta\dot{\nu}_{\rm p}$ and $\Delta\nu$ of small glitches.
Figure \ref{fig5} reveals that the two large events depart greatly from the allowed gray region.
This result strongly denies the external torque origin of persistent shifts due to mass redistribution in starquakes.

As is known to all, changes in magnetosphere could cause radiation changes for the pulsar. However, even for the largest Crab pulsar glitch in 2017, recent dedicated observations following this glitch still revealed no radiation changes. Shaw et al. presented that, by taking use of 42-ft and 76-m Lovell radio telescope at Jordrell Bank Observatory, they found no changes in the pulse profile shape around the glitch epoch at 610~MHz or 1520~MHz, nor did they find any changes in X-ray flux using light curve from the Swift-BAT instrument~\citep{2018MNRAS.478.3832S}. Zhang et al. observed the glitch in the 0.5-10~keV X-ray pulsar Navigation-I satellite to find no variations in the total X-ray flux around the glitch~\citep{2018ApJ...866...82Z}. Vivekanand also found no changes in the soft X-ray spectrum, the flux widths and peaks of its integrated profile following this glitch by analysing the data of the Neutron star Interior Composition Explorer (NICER) satellite~\citep{2020A&A...633A..57V}. Using data collected from Insight-HXMT, NICER and Fermi/GBM, Zhang et al. further constrained the profiles from Insight-HXMT (27-200keV) and NICER (0.5-10keV) around the November 2017 glitch to have on changes with rms 0.47\% and 0.28\%, the pulsed flux remains stable with 1$\sigma$ uncertainty of 0.07\% and 0.011\%~\citep{2022ApJ...932...11Z}.
Our theoretical calculations are consistent with the non-detection of glitch related changes in emission flux, pulse profile, and spectrum of the Crab pulsar.

\section{Discussions and Conclusions}
\label{sec5}

The post-glitch behaviors of pulsars have been seen as chances to test glitch physics.
A successful example of the vortex creep model based on superfluidity is that it faultlessly reproduces the post-glitch relaxation of the Vela and other pulsars in terms of the general parameters of NS structure.
However, understanding the atypical post-glitch behaviors of the Crab pulsar is challenging.
A subject for debate is why glitches cause persistent increases in the spin-down rate of the star but the rotational frequency deficits occur.
A natural thought is for extra torque to be produced through starquakes, independent of the structural readjustments of the star.
In this work, we have disproved two prevailing models that starquakes induce variations of the external torque.
The observational data are inherently inexplicable by these models. The 2017 November glitch is probably a decisive event, which either deviates extremely from the theoretical value of the tectonic model or is contrary to the prediction of the fault model.
In a word, the dependence of persistent shift on glitch size is inconsistent with the external origin of persistent shifts caused by starquakes. Therefore, the magnetosphere will hardly change before and after a glitch for RPPs. The unchanged magnetosphere matches well with the fact that, no radiative changes have been observed around glitches from the Crab pulsar~\citep{2018MNRAS.478.3832S,2018ApJ...866...82Z,2020A&A...633A..57V,2022ApJ...932...11Z}.

If there is no change in external torque of the Crab pulsar around a glitch, we propose an internal origin of persistent shift.
One possibility is that, a small fraction of the superfluid component unpinned from the crustal lattices or flux tubes is destroyed during a glitch, namely, a portion of superfluid matter is converted into normal matter.
Thus, the standard motion equation of the normal component,see Eq.(6a) in~\cite{1984ApJ...276..325A} will be modified to be
\begin{equation}\label{eq11}
    I_{\rm n}\dot{\Omega}=N_{\rm ext}-(I_{\rm s}-\Delta I)\dot{\Omega}_{\rm s},
\end{equation}
where $I_{\rm n}$ and $I_{\rm s}$ denote respectively the moments of inertia coupled with the external torque and the unpinned component, $N_{\rm ext}$ is the external torque of the star, written as $I\dot\Omega_{0,\infty}$, $ I=  I_{\rm n}+I_{\rm s}$ is the total moment of inertia of the star, $\Delta I$ is the moment of inertia converted from superfluid to normal matter.
When the system relax to a new steady state, $ \dot{\Omega}=\dot{\Omega}_{\rm s}=\dot{\Omega}_\infty $, we obtain from Eq.(\ref{eq11}) that, $I\dot{\Omega}_\infty = I\Omega_{0,\infty}+\Delta I\dot{\Omega}_\infty$. Immediately, we have $\Delta\dot{\Omega}_{\rm p}=\dot{\Omega}_\infty-\dot{\Omega}_{0,\infty}={\Delta I\over I}\dot{\Omega}_\infty$.
It implies a persistent shift in the spin-down rate of the pulsar.
Meanwhile, this simple scenario can be allowed for an explanation of the rotational frequency deficit. The total angular moments of neutron stars before and after a glitch read respectively
\begin{equation}\label{eq12}
    J(t)=I\Omega_{0}(t)+I_{\rm s}\omega_0(t)
\end{equation}
\begin{equation}\label{eq13}
    J(t)=I\Omega(t)+I_{\rm s}\omega(t)-\Delta I_{\rm s}\Omega_{\rm s}
\end{equation}
 where $\omega_0(t)$ and $\omega(t)$ represent lags between two components before and after a glitch epoch. Subtract  equation(\ref{eq12}) from (\ref{eq13}) to get $ \Omega(t)- \Omega_{0}(t)={I_{\rm s}\over I}(\omega_0-\omega)-{\Delta I\over I}\Omega_{\rm s}$.
Immediately $ \Omega(t)- \Omega_{0}=-{\Delta I\over I}\Omega_{\rm s}$, always less than zero when two components couple to the steady state, $\omega\rightarrow\omega_0 $. This just exhibits a frequency deficit.
Why are the recovers dramatic differences for the Vela and the Crab pulsar?  We suspect that the Crab pulsar stays around the critical temperature for the $^3$P$_2$ superfluidity~\citep{2012NatPh...8..787H} and then neutron parring is in dynamic instability because  the temperature of the young Crab pulsar is much higher that of the mature Vela pulsar.

In conclusions, our study presented the glitch dependence of spin-down shift instead of individual case to understand the post-glitch behavior of the Crab pulsar. The research findings support no star-quake on the star, meeting with no observed changes in the emission flux, pulse profile and spectrum of the Crab pulsar. We also give  some estimates of internal process. If this were the real case, the glitch behavior of the Crab pulsar is a more valuable probe than the Vela pulsar for nucleonic superfluidity in neutron stars, worth exploring further. For this sake, we need detailed investigations of dynamics to fit the glitch-size dependence of the persistent shifts in future.

\section*{acknowledgements} This paper is supported  by the National SKA Program of China (Grant No. 2020SKA0120300) and the National Natural Science Foundation of China (Grant Nos. 12033001,12473039). Weihua Wang is supported by Zhejiang Provincial Natural Science Foundation of China under grant No. LQ24A030002.

\bibliography{persistent_shift}{}

\begin{thebibliography}{}
\expandafter\ifx\csname natexlab\endcsname\relax\def\natexlab#1{#1}\fi
\providecommand{\url}[1]{\href{#1}{#1}}
\providecommand{\dodoi}[1]{doi:~\href{http://doi.org/#1}{\nolinkurl{#1}}}
\providecommand{\doeprint}[1]{\href{http://ascl.net/#1}{\nolinkurl{http://ascl.net/#1}}}
\providecommand{\doarXiv}[1]{\href{https://arxiv.org/abs/#1}{\nolinkurl{https://arxiv.org/abs/#1}}}

\bibitem[{{Akbal} \& {Alpar}(2018)}]{2018MNRAS.473..621A}
{Akbal}, O., \& {Alpar}, M.~A. 2018, \mnras, 473, 621,
  \dodoi{10.1093/mnras/stx2378}

\bibitem[{{Akbal} {et~al.}(2015){Akbal}, {G{\"u}gercino{\u{g}}lu},
  {{\c{S}}a{\c{s}}maz Mu{\c{s}}}, \& {Alpar}}]{2015MNRAS.449..933A}
{Akbal}, O., {G{\"u}gercino{\u{g}}lu}, E., {{\c{S}}a{\c{s}}maz Mu{\c{s}}}, S.,
  \& {Alpar}, M.~A. 2015, \mnras, 449, 933, \dodoi{10.1093/mnras/stv322}

\bibitem[{{Alpar} {et~al.}(1993){Alpar}, {Chau}, {Cheng}, \&
  {Pines}}]{1993ApJ...409..345A}
{Alpar}, M.~A., {Chau}, H.~F., {Cheng}, K.~S., \& {Pines}, D. 1993, \apj, 409,
  345, \dodoi{10.1086/172668}

\bibitem[{{Alpar} {et~al.}(1996){Alpar}, {Chau}, {Cheng}, \&
  {Pines}}]{1996ApJ...459..706A}
---. 1996, \apj, 459, 706, \dodoi{10.1086/176935}

\bibitem[{{Alpar} {et~al.}(1984){Alpar}, {Pines}, {Anderson}, \&
  {Shaham}}]{1984ApJ...276..325A}
{Alpar}, M.~A., {Pines}, D., {Anderson}, P.~W., \& {Shaham}, J. 1984, \apj,
  276, 325, \dodoi{10.1086/161616}

\bibitem[{{Anderson} \& {Itoh}(1975)}]{1975Natur.256...25A}
{Anderson}, P.~W., \& {Itoh}, N. 1975, \nat, 256, 25, \dodoi{10.1038/256025a0}

\bibitem[{{Basu} {et~al.}(2020){Basu}, {Joshi}, {Krishnakumar}, {Bhattacharya},
  {Nandi}, {Bandhopadhay}, {Char}, \& {Manoharan}}]{2020MNRAS.491.3182B}
{Basu}, A., {Joshi}, B.~C., {Krishnakumar}, M.~A., {et~al.} 2020, \mnras, 491,
  3182, \dodoi{10.1093/mnras/stz3230}

\bibitem[{{Basu} {et~al.}(2022){Basu}, {Shaw}, {Antonopoulou}, {Keith}, {Lyne},
  {Mickaliger}, {Stappers}, {Weltevrede}, \& {Jordan}}]{2022MNRAS.510.4049B}
{Basu}, A., {Shaw}, B., {Antonopoulou}, D., {et~al.} 2022, \mnras, 510, 4049,
  \dodoi{10.1093/mnras/stab3336}

\bibitem[{{Baym} {et~al.}(1969){Baym}, {Pethick}, \&
  {Pines}}]{1969Natur.224..673B}
{Baym}, G., {Pethick}, C., \& {Pines}, D. 1969, \nat, 224, 673,
  \dodoi{10.1038/224673a0}

\bibitem[{{Baym} \& {Pines}(1971)}]{1971AnPhy..66..816B}
{Baym}, G., \& {Pines}, D. 1971, Annals of Physics, 66, 816,
  \dodoi{10.1016/0003-4916(71)90084-4}

\bibitem[{{Epstein} \& {Link}(2000)}]{2000ASSL..254...95E}
{Epstein}, R.~I., \& {Link}, B. 2000, in Astrophysics and Space Science
  Library, Vol. 254, Astrophysics and Space Science Library, ed. K.~S. {Cheng},
  H.~F. {Chau}, K.~L. {Chan}, \& K.~C. {Leung}, 95,
  \dodoi{10.1007/978-94-010-0878-5_12}

\bibitem[{{Espinoza} {et~al.}(2014){Espinoza}, {Antonopoulou}, {Stappers},
  {Watts}, \& {Lyne}}]{2014MNRAS.440.2755E}
{Espinoza}, C.~M., {Antonopoulou}, D., {Stappers}, B.~W., {Watts}, A., \&
  {Lyne}, A.~G. 2014, \mnras, 440, 2755, \dodoi{10.1093/mnras/stu395}

\bibitem[{{Flanagan}(1990)}]{1990Natur.345..416F}
{Flanagan}, C.~S. 1990, \nat, 345, 416, \dodoi{10.1038/345416a0}

\bibitem[{{Franco} {et~al.}(2000){Franco}, {Link}, \&
  {Epstein}}]{2000ApJ...543..987F}
{Franco}, L.~M., {Link}, B., \& {Epstein}, R.~I. 2000, \apj, 543, 987,
  \dodoi{10.1086/317121}

\bibitem[{{Giliberti} {et~al.}(2020){Giliberti}, {Cambiotti}, {Antonelli}, \&
  {Pizzochero}}]{2020MNRAS.491.1064G}
{Giliberti}, E., {Cambiotti}, G., {Antonelli}, M., \& {Pizzochero}, P.~M. 2020,
  \mnras, 491, 1064, \dodoi{10.1093/mnras/stz3099}

\bibitem[{{Gullahorn} {et~al.}(1977){Gullahorn}, {Isaacman}, {Rankin}, \&
  {Payne}}]{1977AJ.....82..309G}
{Gullahorn}, G.~E., {Isaacman}, R., {Rankin}, J.~M., \& {Payne}, R.~R. 1977,
  \aj, 82, 309, \dodoi{10.1086/112050}

\bibitem[{{Ho} \& {Andersson}(2012)}]{2012NatPh...8..787H}
{Ho}, W. C.~G., \& {Andersson}, N. 2012, Nature Physics, 8, 787,
  \dodoi{10.1038/nphys2424}

\bibitem[{{Ho} {et~al.}(2022){Ho}, {Kuiper}, {Espinoza}, {Guillot}, {Ray},
  {Smith}, {Bogdanov}, {Antonopoulou}, {Arzoumanian}, {Bejger}, {Enoto},
  {Esposito}, {Harding}, {Haskell}, {Lewandowska}, {Maitra}, \&
  {Vasilopoulos}}]{2022ApJ...939....7H}
{Ho}, W. C.~G., {Kuiper}, L., {Espinoza}, C.~M., {et~al.} 2022, \apj, 939, 7,
  \dodoi{10.3847/1538-4357/ac8743}

\bibitem[{{Hobbs} {et~al.}(2006){Hobbs}, {Edwards}, \&
  {Manchester}}]{2006MNRAS.369..655H}
{Hobbs}, G.~B., {Edwards}, R.~T., \& {Manchester}, R.~N. 2006, \mnras, 369,
  655, \dodoi{10.1111/j.1365-2966.2006.10302.x}

\bibitem[{{Kou} {et~al.}(2018){Kou}, {Yuan}, {Wang}, {Yan}, \&
  {Dang}}]{2018MNRAS.478L..24K}
{Kou}, F.~F., {Yuan}, J.~P., {Wang}, N., {Yan}, W.~M., \& {Dang}, S.~J. 2018,
  \mnras, 478, L24, \dodoi{10.1093/mnrasl/sly068}

\bibitem[{{Lai} {et~al.}(2023){Lai}, {Wang}, {Yuan}, {Lu}, {Yue}, \&
  {Xu}}]{2023MNRAS.523.3967L}
{Lai}, X.~Y., {Wang}, W.~H., {Yuan}, J.~P., {et~al.} 2023, \mnras, 523, 3967,
  \dodoi{10.1093/mnras/stad1653}

\bibitem[{{Lai} {et~al.}(2018){Lai}, {Yun}, {Lu}, {L{\"u}}, {Wang}, \&
  {Xu}}]{2018MNRAS.476.3303L}
{Lai}, X.~Y., {Yun}, C.~A., {Lu}, J.~G., {et~al.} 2018, \mnras, 476, 3303,
  \dodoi{10.1093/mnras/sty474}

\bibitem[{{Link} \& {Epstein}(1997)}]{1997ApJ...478L..91L}
{Link}, B., \& {Epstein}, R.~I. 1997, \apjl, 478, L91, \dodoi{10.1086/310549}

\bibitem[{{Link} {et~al.}(1992){Link}, {Epstein}, \&
  {Baym}}]{1992ApJ...390L..21L}
{Link}, B., {Epstein}, R.~I., \& {Baym}, G. 1992, \apjl, 390, L21,
  \dodoi{10.1086/186362}

\bibitem[{{Link} {et~al.}(1999){Link}, {Epstein}, \&
  {Lattimer}}]{1999PhRvL..83.3362L}
{Link}, B., {Epstein}, R.~I., \& {Lattimer}, J.~M. 1999, \prl, 83, 3362,
  \dodoi{10.1103/PhysRevLett.83.3362}

\bibitem[{{Link} {et~al.}(1998){Link}, {Franco}, \&
  {Epstein}}]{1998ApJ...508..838L}
{Link}, B., {Franco}, L.~M., \& {Epstein}, R.~I. 1998, \apj, 508, 838,
  \dodoi{10.1086/306457}

\bibitem[{{Lohsen}(1981)}]{1981A&AS...44....1L}
{Lohsen}, E.~H.~G. 1981, \aaps, 44, 1

\bibitem[{{Lyne} {et~al.}(2015){Lyne}, {Jordan}, {Graham-Smith}, {Espinoza},
  {Stappers}, \& {Weltevrede}}]{2015MNRAS.446..857L}
{Lyne}, A.~G., {Jordan}, C.~A., {Graham-Smith}, F., {et~al.} 2015, \mnras, 446,
  857, \dodoi{10.1093/mnras/stu2118}

\bibitem[{{Manchester}(2018)}]{2018IAUS..337..197M}
{Manchester}, R.~N. 2018, in IAU Symposium, Vol. 337, Pulsar Astrophysics the
  Next Fifty Years, ed. P.~{Weltevrede}, B.~B.~P. {Perera}, L.~L. {Preston}, \&
  S.~{Sanidas}, 197--202, \dodoi{10.1017/S1743921317009607}

\bibitem[{{Ng} {et~al.}(2016){Ng}, {Takata}, \& {Cheng}}]{2016ApJ...825...18N}
{Ng}, C.~W., {Takata}, J., \& {Cheng}, K.~S. 2016, \apj, 825, 18,
  \dodoi{10.3847/0004-637X/825/1/18}

\bibitem[{{Palfreyman} {et~al.}(2018){Palfreyman}, {Dickey}, {Hotan},
  {Ellingsen}, \& {van Straten}}]{2018Natur.556..219P}
{Palfreyman}, J., {Dickey}, J.~M., {Hotan}, A., {Ellingsen}, S., \& {van
  Straten}, W. 2018, \nat, 556, 219, \dodoi{10.1038/s41586-018-0001-x}

\bibitem[{{Pines} \& {Shaham}(1972)}]{1972NPhS..235...43P}
{Pines}, D., \& {Shaham}, J. 1972, Nature Physical Science, 235, 43,
  \dodoi{10.1038/physci235043a0}

\bibitem[{{Rencoret} {et~al.}(2021){Rencoret}, {Aguilera-G{\'o}mez}, \&
  {Reisenegger}}]{2021A&A...654A..47R}
{Rencoret}, J.~A., {Aguilera-G{\'o}mez}, C., \& {Reisenegger}, A. 2021, \aap,
  654, A47, \dodoi{10.1051/0004-6361/202141499}

\bibitem[{{Ruderman}(1969)}]{1969Natur.223..597R}
{Ruderman}, M. 1969, \nat, 223, 597, \dodoi{10.1038/223597b0}

\bibitem[{{Ruderman}(1991)}]{1991ApJ...366..261R}
---. 1991, \apj, 366, 261, \dodoi{10.1086/169558}

\bibitem[{{Shaw} {et~al.}(2021){Shaw}, {Keith}, {Lyne}, {Mickaliger},
  {Stappers}, {Turner}, \& {Weltevrede}}]{2021MNRAS.505L...6S}
{Shaw}, B., {Keith}, M.~J., {Lyne}, A.~G., {et~al.} 2021, \mnras, 505, L6,
  \dodoi{10.1093/mnrasl/slab038}

\bibitem[{{Shaw} {et~al.}(2018){Shaw}, {Lyne}, {Stappers}, {Weltevrede},
  {Bassa}, {Lien}, {Mickaliger}, {Breton}, {Jordan}, {Keith}, \&
  {Krimm}}]{2018MNRAS.478.3832S}
{Shaw}, B., {Lyne}, A.~G., {Stappers}, B.~W., {et~al.} 2018, \mnras, 478, 3832,
  \dodoi{10.1093/mnras/sty1294}

\bibitem[{{Vivekanand}(2017)}]{2017A&A...597L...9V}
{Vivekanand}, M. 2017, \aap, 597, L9, \dodoi{10.1051/0004-6361/201630235}

\bibitem[{{Vivekanand}(2020)}]{2020A&A...633A..57V}
---. 2020, \aap, 633, A57, \dodoi{10.1051/0004-6361/201936774}

\bibitem[{{Wang} {et~al.}(2021){Wang}, {Lai}, {Zhou}, {Lu}, {Zheng}, \&
  {Xu}}]{2021MNRAS.500.5336W}
{Wang}, W.~H., {Lai}, X.~Y., {Zhou}, E.~P., {et~al.} 2021, \mnras, 500, 5336,
  \dodoi{10.1093/mnras/staa3520}

\bibitem[{{Zhang} {et~al.}(2018){Zhang}, {Shuai}, {Huang}, {Chen}, \&
  {Du}}]{2018ApJ...866...82Z}
{Zhang}, X., {Shuai}, P., {Huang}, L., {Chen}, S., \& {Du}, Y. 2018, \apj, 866,
  82, \dodoi{10.3847/1538-4357/aade46}

\bibitem[{{Zhang} {et~al.}(2022){Zhang}, {Ge}, {Lu}, {Tuo}, {Song}, {Zhang},
  {Wang}, {Zheng}, \& {Yan}}]{2022ApJ...932...11Z}
{Zhang}, Y.~H., {Ge}, M.~Y., {Lu}, F.~J., {et~al.} 2022, \apj, 932, 11,
  \dodoi{10.3847/1538-4357/ac6d53}

\bibitem[{{Zhou} {et~al.}(2014){Zhou}, {Lu}, {Tong}, \&
  {Xu}}]{2014MNRAS.443.2705Z}
{Zhou}, E.~P., {Lu}, J.~G., {Tong}, H., \& {Xu}, R.~X. 2014, \mnras, 443, 2705,
  \dodoi{10.1093/mnras/stu1370}

\bibitem[{{Zubieta} {et~al.}(2024){Zubieta}, {Garc{\'\i}a}, {del Palacio},
  {Espinoza}, {Araujo Furlan}, {Gancio}, {Lousto}, {Combi}, \&
  {G{\"u}gercino{\u{g}}lu}}]{2024arXiv241217766Z}
{Zubieta}, E., {Garc{\'\i}a}, F., {del Palacio}, S., {et~al.} 2024, arXiv
  e-prints, arXiv:2412.17766, \dodoi{10.48550/arXiv.2412.17766}

\end{thebibliography}
\bibliographystyle{aasjournal}

\end{document}